\documentclass[%
 reprint,
amsmath,amssymb,
 aps
]{revtex4-1}

\usepackage{graphicx}
\usepackage{dcolumn}
\usepackage{bm}
\usepackage{epstopdf}
\usepackage{textcomp}
\usepackage{gensymb}
\usepackage{amssymb}
\usepackage{amsmath}
\usepackage{ upgreek }
\usepackage{esvect}

\begin{document}
\title{Synthetically Encapsulated \& Self-Organized Transition Metal Oxide Nano Structures inside Carbon Nanotubes as Robust Li-ion Battery Anode Materials}
\author{Aakanksha Kapoor$^1$, Apurva L.Patrike$^1$, Nitesh Singh$^{1}$, Elisa Thauer$^2$, Alexander Ottmann$^2$, R\"{u}diger Klingeler$^{2,3}$, Satishchandra Ogale$^{1,4}$, and A. Bajpai$^{1,5}$}
\affiliation{$^1 $Department of Physics, Indian Institute of Science Education and Research, Pune - 411008, India}
\affiliation{$^2$Kirchhoff Institute of Physics, Heidelberg University, D-69120 Heidelberg, Germany}
\affiliation{$^3 $Centre for Advanced Materials (CAM), Heidelberg University, D-69120 Heidelberg, Germany}
\affiliation{$^4$Research Institute for Sustainable Energy (RISE), TCG-Centres for Research and Education in Science and Technology, Sector V, Salt Lake, Kolkata - 700091, India}
\affiliation{$^5 $Centre for Energy Science, Indian Institute of Science Education and Research, Pune - 411008, India}
\date{\today}
\email[]{ashna@iiserpune.ac.in}

\begin{abstract}

We report a comprehensive study on the electrochemical performance of four different Transition Metal Oxides encapsulated inside carbon nanotubes (CNT). Irrespective of the type of oxide-encapsulate, all these samples exhibit superior cyclic stability as compared to the bare-oxide.  Innovative use of camphor during sample synthesis enables precise control over the morphology of these self-organized carbon nanotube structures, which in turn appears to play a crucial role in the magnitude of the specific capacity. A comparative evaluation of the electrochemical data on different samples bring forward interesting inferences pertaining to the morphology, filling fraction of the oxide-encapsulate, and the presence of oxide nano-particles adhering outside the filled CNT. Our results provides useful pointers towards the optimization of critical parameters, thus paving the way for using these synthetically encapsulated and self-organized carbon nanotube structures  as anode materials for Li-ion batteries, and possibly other electrochemical applications.	
		
\end{abstract}

\pacs{Valid PACS appear here}
\maketitle


\section{\label{sec:level1}Introduction}

Li-ion batteries (LIBs) are a critical component of most advanced technologies of the modern era, and represent perhaps the most sought after energy storage devices of these times.\cite{Goodenough, Peters, Hameer, Aravindan} The commercial LIBs employ graphite as the anode material which has a theoretical capacity of 372 mA h g$^{-1}$.\cite{Ji} Transition metal oxides (TMOs) have attracted considerable attention as promising anode materials due to their higher theoretical capacities in comparison to graphite. Additional advantages of these TMOs include low cost, abundant resources, and low toxicity. Though the experimentally observed capacities are lower, these can be substantially improved via nano scaling. Therefore, TMOs have been extensively explored in their nano-flower, nano-flake, and nano-porous morphologies.\cite{ Zhang,Zheng,Tian,Cao, Zhang1,Fan,Liu,Reddy,Li,Pang} The lithium storage capacity of TMOs is ascribed to the reversible reaction between Li ions and metal oxide(s) which leads to the formation of metal nanoparticles in a Li$_2$O matrix. Superior capacity of TMOs can also be attributed to its storage mechanism, in which the number of Li ions available for intercalation is more compared to the classical intercalation, limited to 1 Li$^+$/f.u.\cite{ Zhang,Zheng,Tian,Cao, Zhang1,Fan,Liu,Reddy,Li,Pang} However, the practical applications using TMOs as an active material are primarily hampered due to the large volume expansion and contraction during the metal to metal-oxide conversion. This conversion leads to a rupture of the separator and destruction of the electrode during the electrochemical reaction.\cite{Ji} This issue of capacity fading which adversely affects the cyclic stability, along with the intrinsic low electrical conductivity are major detrimental factors for TMOs based LIBs.

An effective strategy to improve the structural stability and the electrical conductivity is the use of hybrids based on carbonaceous materials and TMOs in the electrode design.\cite{Peters,Song,Ji,Wang,Wua,Zhao,Jiang,Yang,Wang1,Xu} In this context, CNT/ TMOs hybrids have been explored, as CNT possess excellent electrical conductivity and mechanical stability\cite{Yan,Zhou,Zhang2,Cao1,Alexander,Ogale,Tang,Liu1,Otto,Eliza,Choi}. The exceptional mechanical strength of CNT can also be utilized to buffer the volumetric strain in the electrode materials during charging/discharging processes\cite{Yan,Zhou,Zhang2,Cao1,Alexander,Ogale,Tang,Liu1, Xu,Choi,Otto,Eliza}. Here it is important to distinguish two distinct scenarios pertaining to the usage of oxide-CNT hybrids for LIB. In the first, the CNT and the oxide are synthesized separately and processed to form the composite \cite{Cycling1,Xu,Song,Wang1,Yang,Cao1,Choi,Wang2,Cao3}. Alternatively, the oxide is synthetically encapsulated within the core cavity of the CNT in the form of nano particles or nano wires \cite{Yan,Alexander,Ogale,Liu1}. The core cavity of the CNT in this case can provide space for the volume-changes that take place during metal to metal-oxide conversion. In this work, we present extensive electrochemical data on self-organized  CNT structures, pertaining to the second case, with the TMOs being synthetically encapsulated within the core cavity of the CNT (Oxides@CNT). We reiterate that while FeO$_x$@CNT has been tested as an anode material,\cite{Wang2,Liu1,Yan,Cheng,Liu2} to the best of our knowledge, CoO$_x$ \& NiO$_x$@CNT with significant filling efficiencies (such as presented in this work) have not yet been explored for LIBs, primarily due to difficulties related to the synthesis of such systems \cite{Kapoor}.

\begin{figure}[!t]
	\centering
	\includegraphics[width=0.4\textwidth, height=0.4\textwidth]{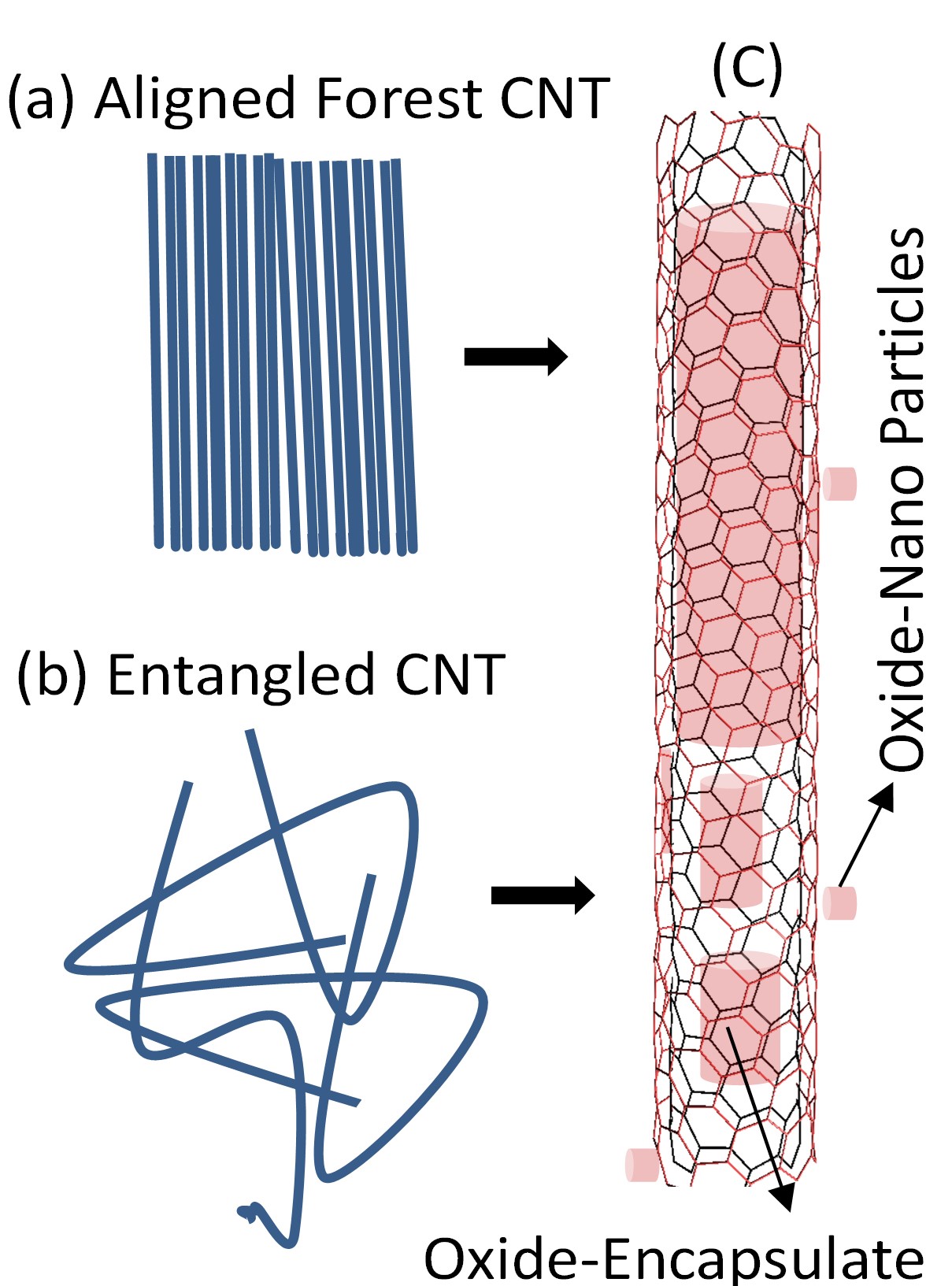}
		\caption{\textit{(a-b) display a schematic for \textit{aligned-forest} and \textit{entangled} morphologies for Oxides@CNT. (c) shows the enlarged schematic of an individual (multi-walled) CNT, which encapsulates oxide in the form of long or short nano-wires within its core cavity. Residue oxide nano particles adhering outside CNT shown in the schematic exist in both type of morphologies.}}
	\label{fig:1}
\end{figure}

In addition to the morphology of the active material, other important factors that are likely to influence the overall battery performance for Oxides@CNT are (i) type of the oxide encapsulate and its size and shape, (ii) filling efficiencies, and (iii) the role played by oxide nano-particles (NP) adhering outside the CNT.\cite{Ji,Otto} In addition to such residual NP, there can be big chunks of oxide clusters, which at times form during the synthesis. Here we present comparative electrochemical data on four different types of TMOs encapsulated inside CNT. Our key result is that the encapsulation of TMOs within the core cavity of CNT leads to a significant improvement in the cyclic stability of the electrode, irrespective of its type. The paper is organized as follows. We first present complete structural and electrochemical characterization, bringing out the correlations between the two. For this, Fe$_{3}$O$_{4}$@CNT, Co$_{3}$O$_{4}$@CNT and NiO@CNT are investigated. The effect of morphology is further explored with focus on Fe$_{2}$O$_{3}$@CNT. These data reveal how the over-all morphology as well as oxide NP residing outside filled CNTs affect the cyclic stability. Morphology variations in Fe$_{2}$O$_{3}$@CNT also establish that Oxides@CNT formed in \textit{entangled} morphology is preferable for the cyclic stability and may play a role in improving the magnitude of capacity. Finally, we compare the cyclic stability of different Oxides@CNT with that of a representative bare-oxide. Here the bare-oxide has been derived from the corresponding Oxide@CNT, and hence retains the same morphology. This comparison enables us to isolate the advantages associated with the encapsulation of TMOs inside CNT.

\section{Experimental Section}

\subsection{Oxides@CNT : Synthesis and Characterization }
The furnace employed for synthesis of the oxides filled CNT was Nabertherm R 100/750/13.The Scanning Electron Microscopy (FESEM) images were recorded using ZEISS ULTRA plus field-emission FESEM for morphological studies. All the samples have been characterized using X-ray powder diffraction (XRD) using Bruker D8 Advance with Cu-K$\alpha$ radiation ($\lambda$ = 1.54056\AA). The thermal analysis was determined by a thermogravimetric analyser, Perkin Elmer STA 6000, under air at 20 mL min$^{-1}$ at a heating rate of 15$^o$C min$^{-1}$ from 30$^o$C to 900$^o$C. Raman spectroscopy measurements were performed on HORIBA JOBIN YVON LabRam HR 800 with an excitation wavelength of 488nm.

The high quality samples of oxides@CNT are synthesized by using a two step process. The first step involves preparation of  metal@CNT (Fe@CNT, Co@CNT and Ni@CNT) by using pyrolysis of metallocene.\cite{Kapoor} A few representative TEM images of all three metal@CNT are shown in Supp. Information: \textbf{Figure}\textbf{ S1(a-c)} for Fe@CNT, \textbf{S2(a-c)} for Ni@CNT, and \textbf{S3(a-c)} for Co@CNT. In the second step, the metal@CNT are converted to their respective Oxide@CNT using a suitable annealing protocol \cite{Kapoor}. Broad area FESEM images, depicting overall morphology of  various Oxides@CNT are shown in Supp. Information: \textbf{Figure S1(d-g)} for FeO$_x$@CNT. In this context, we note that FeO$_x$@CNT has been tested earlier for Li-ion batteries.\cite{Wang2,Liu1,Yan,Cheng,Liu2} However,  Ni@CNT and Co@CNT, with well formed graphitic shells, along with substantial filling efficiencies are difficult to form by pyrolysis of metallocene, a routine synthesis technique for the formation of Fe@CNT\cite{Kapoor}. The co- pyrolysis of metallocene  with the cost-effective and eco-friendly compound \textit{camphor}, has enabled the formation of well formed samples of  Ni@CNT and Co@CNT\cite{Kapoor}. This in turn enables well formed  samples of  NiO@CNT and CoO$_x$@CNT, as evident from FESEM images shown in Sup. Info: \textbf{Figures S2} \&  \textbf{S3}. The usage of camphor also provides an additional tool to tailor the morphology, which is an important factor for LIB applications.  For each type of oxide to form in a certain morphology, the ratio of metallocene and camphor can be optimized with other experimental parameters associated with pyrolysis of metallocene\cite{Kapoor}. The metal@CNT (and their corresponding oxide@CNT) presented in this work can be  broadly classified to form either in  (i) \textit{aligned-forest} morphology, or (ii) \textit{entangled} morphology. The bare oxide Fe$_{2}$O$_{3}$ has been obtained in the same morphology as Fe$_{2}$O$_{3}$@CNT. This is achieved by suitably annealing the Fe$_{2}$O$_{3}$@CNT sample\cite{Kapoor2}. The integrity of this oxide-template (bare-oxide) is checked through XRD and FESEM.\cite{Kapoor2} 
 
\subsection{Electrochemical Measurements Protocol}
The electrochemical behavior of Oxides@CNT was studied using CR2032 coin type and Swagelok-type cells. The Oxide@CNT powders were sonicated for 30 minutes prior the electrode preparation in both the cases. For CR2032 coin type cells using  lithium foil as the counter and reference electrode, a Whatman membrane was used as a separator, and 1M LiPF$_{6}$ in ethylene carbonate and diethyl carbonate (EC:DEC = 1:1 v/v) was used as the electrolyte. The working electrode was fabricated by compressing a mixture of the active material (Oxides@CNT), a conductive material (acetylene black), and a binder (polyvinylidene fluoride) in a weight ratio of $8:1:1$ onto a copper foil current collector. This  was kept at 80 $^o$C for 12 hours and the cell  was assembled in an argon-filled glove box. The cells were galvanostatically charged and discharged in the voltage range of 0.01 V to 3.0 V. All the electrochemical measurements are performed on a MTI corporation  battery analyzer. Cyclic voltammetry was performed using an Ametek potentiostat at a scan-rate of 0.1  mV s$^{-1}$ from 0.01 to 3 V, with the impedance spectra being measured in the frequency range of 10 mHz - 300 kHz.  
 
\begin{figure*}[!t]
	\centering
	\includegraphics[width=1\textwidth]{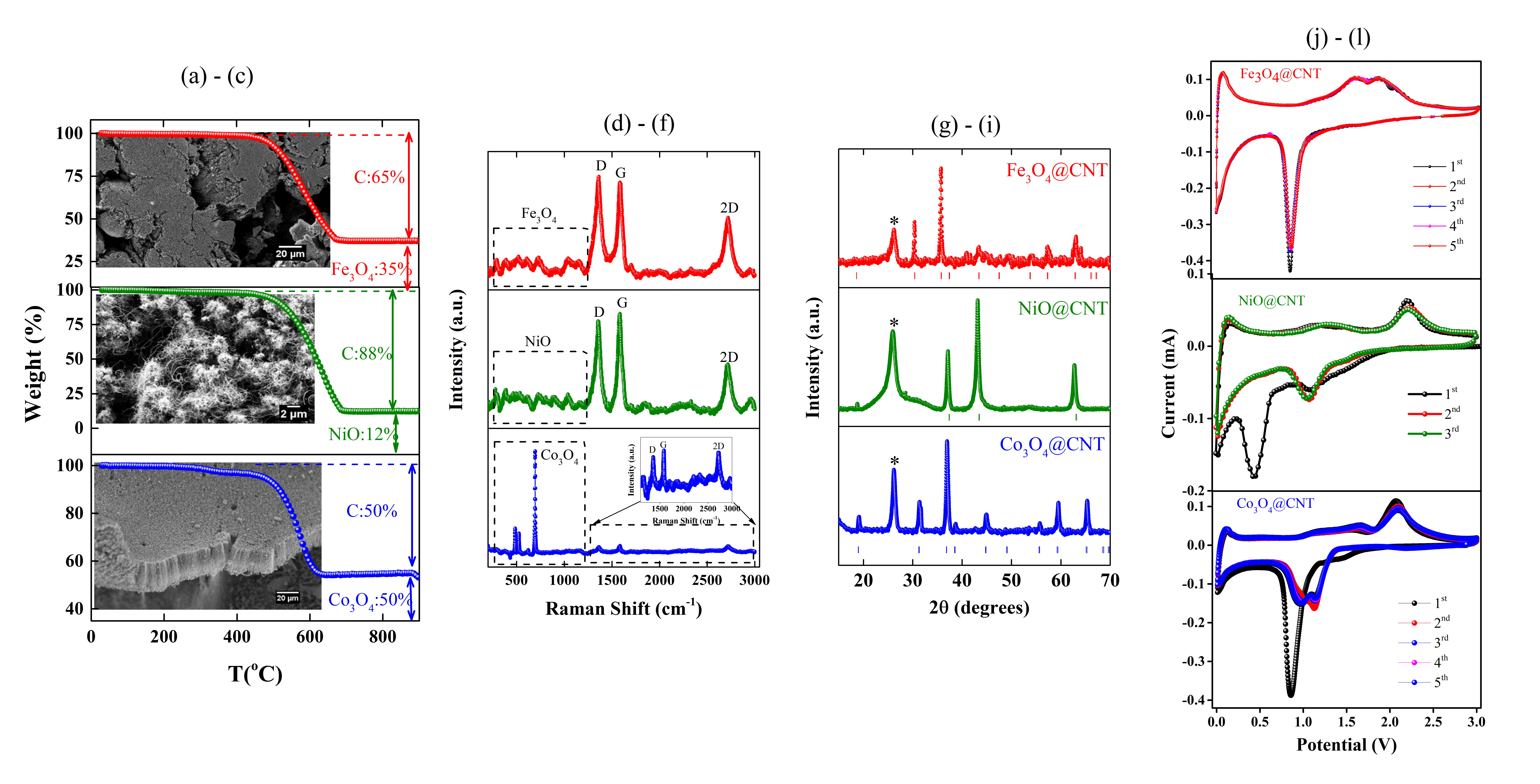}
	\caption{ \textit{(a-c) displays TGA data for Fe$_{3}$O$_{4}$@CNT (red dots), NiO@CNT (green dots) and Co$_{3}$O$_{4}$@CNT (blue dots). The carbon and the oxide content as obtained from TGA is indicated in (a-c). Broad area FESEM images in the respective insets reveal that Fe$_{3}$O$_{4}$@CNT as well as Co$_{3}$O$_{4}$@CNT are formed in the \textit{aligned-forests} morphology and the NiO@CNT is formed in \textit{entangled} morphology. (d-f) shows Raman spectra depicting D, G, and 2D bands corresponding to CNT. The characteristic peaks corresponding to the oxide encapsulate is also identified. (g-i) shows XRD data, with star in each panel denoting the Bragg peak corresponding to CNT and the remaining peaks are identified with respective oxides. (j-l) shows Cyclic Voltammetry (CV) curves of Fe$_{3}$O$_{4}$@CNT, NiO@CNT and Co$_{3}$O$_{4}$@CNT, between 3.0 and 0.1 V at a scan rate of 0.1 mV s$^{-1}$, characterizing oxide-metal conversion for each sample.}}	\label{fig:2}
\end{figure*}

The materials were also tested using Swagelok-type cells \cite{Otto1} using a VMP3 potentiostat (Bio-Logic SAS) at a temperature of 25 $^o$C. For the preparation of the working electrode, polyvinylidene fluoride (PVDF, Solvay Plastics) was dissolved in NMP (N-methyl-2-pyrrolidone). Active material as well as carbon black (Super C65, Timcal) were added with mass ratio 1:8:1 and stirred for 12 hours. The resulting mixture was dried in a vacuum oven (65$^o$C, p $<$ 10mbar) until a spreadable slurry was formed, which was then applied on a copper mesh current collectors (10 mm diameter). The as-prepared electrodes were dried at 80 $^o$C in a vacuum oven ($<$ 10 mbar), mechanically pressed at 10 MPa, and dried again. The cells were assembled in a glovebox under argon atmosphere (O$_2$/H$_2$O $<$ 5 ppm) using a lithium metal foil disk (Alfa Aesar) pressed on a nickel current collector (diameter 12 mm) as counter electrode. The electrodes were separated by two layers of glass microfibre (Whatman GF/D) soaked with 200 $\mu$l of a 1 M LiPF$_6$ salt solution in 1:1 ethylene carbonate and dimethyl carbonate (Merck Electrolyte LP30). 

\section{Results and Discussion}

\subsection{ Morphology of Oxides@CNT : Aligned-Forest and Entangled}
We first present a schematic for defining the two types of morphology of Oxides@CNT used in this work. First is the \textit{aligned-forest} depicted in \textbf{Figure 1(a)}. Here the CNT network forms large \textit{carpet-like} structures. The second is \textit{entangled} morphology as shown in \textbf{Figure 1(b)}, which consists of individual curled CNTs. Individual CNTs in both these morphologies are open, multi-walled, and contains oxide-filling in the form of long and short nano-wires, as depicted schematically in \textbf{Figure 1(c)}. In both these morphologies, the oxide NPs can adhere to the outermost wall of the CNT (\textbf{Figure 1(c)}). The density of such oxide NPs in real samples is seen to vary, depending on the synthesis conditions.\cite{Kapoor} Oxides@CNT formed in \textit{aligned-forest} morphology, such as shown in \textbf{Figure 1(a)}, lead to a narrow distribution of length and diameter of the individual CNT, which can be favorable for the reproducibility of the electrochemical data. However, Oxides@CNT in \textit{entangled} morphology (\textbf{Figure 1(b)}) is likely to be more conducive for Li-ion transportation during the electrochemical process, owing to the additional space available for the intercalation of Li ions. The morphology of the Oxides@CNT also controls the conductance of the CNT network, and therefore relates to the overall performance of the cell, including capacity as well as cyclic stability. Therefore, correlations between the structural and electrochemical characterization can enable the optimization of best parameters for Oxides@CNT systems for LIB.      
     
\subsection {Fe$_{3}$O$_{4}$@CNT, Co$_{3}$O$_{4}$ and NiO@CNT: Structural Characterization}
\textbf{Figure 2(a-c)} shows the results of TGA measurements (main panel) along with broad area FESEM images (inset) for all three types of Oxides@CNT. The TGA data reveal the amount of carbon in Oxides@CNT, through the weight loss during the process of heating. The CNT and the oxide ratio, as observed from respective TGA data is indicated in the main panel of \textbf{Figure 2}. The oxide weight content was estimated to be 35\% in the case of Fe$_{3}$O$_{4}$@CNT (red dots), 12\% for NiO@CNT (green dots), and 54\% in case of Co$_{3}$O$_{4}$@CNT (blue dots). While these data can give a rough estimate of the ratio of carbon and oxide in each sample, such measurements cannot confirm whether the oxide is inside CNT (in the form of encapsulate within its core cavity) or adhering outside the CNT (in the form of NP). The estimate for filling efficiencies and residual NP density can be further refined using FESEM and TEM images.\cite{Kapoor,Ashna,Peci}

\begin{figure*}[!t]
	\centering
	\includegraphics[width=1\textwidth]{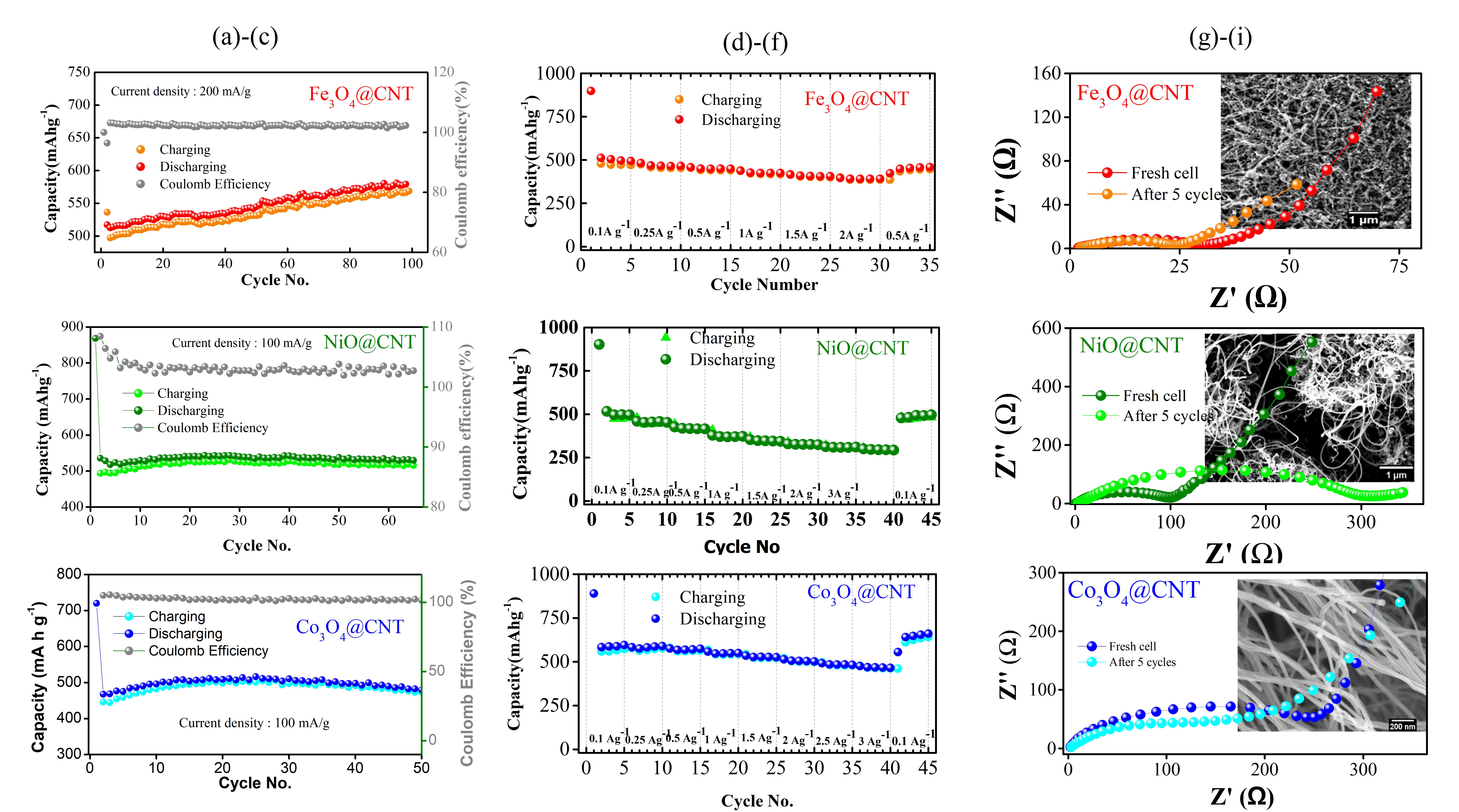}
	\caption{ \textit{(a)-(c) shows discharge and charge capacities and Coulomb efficiencies for Fe$_{3}$O$_{4}$@CNT (red dots), NiO@CNT (green dots) and Co$_{3}$O$_{4}$@CNT (blue dots), as a function of the cycle number. Discharge and charge capacity of Oxides@CNT at different current densities are shown in (d)-(f). (g-i) shows electrochemical impedance spectroscopy (EIS) spectra for fresh cell (black dots) and post five cycles (colored dots) for all three samples. The inset in each panel shows high magnification FESEM image of respective Oxides@CNT.}} 
	\label{fig:3}
\end{figure*}

Broad area FESEM images, representative ones shown in the inset of \textbf{Figure 2(a-c)}, reveal the overall morphology for each type of sample. Multiple FESEM and TEM images (Suppl. Info. \textbf{Figure S1-S3}) not only confirm the presence of the encapsulate, but also provides an estimation of residual NP density.\cite{Kapoor} From FESEM images shown in \textbf{Figure 2(a-c)}, it is also evident that Fe$_{3}$O$_{4}$@CNT and Co$_{3}$O$_{4}$@CNT have been formed in carpet-like structures, with area in the range of $\sim$ 100 square $\mu$m. These carpets have typical thickness $\sim$ 20-30 $\mu$m. On closer inspection, carpets in the case of Fe$_{3}$O$_{4}$@CNT consist of individual CNT which are \textit{entangled}, a morphology similar to what is shown schematically in \textbf{Figure 1(b)}. The length of individual CNT in this case varies from 100-500 nm. On the other hand, the carpets in the case of Co$_{3}$O$_{4}$@CNT consist of \textit{aligned-forest} morphology (\textbf{Figure 1(a)}). The typical length and the outer diameter of the individual CNT in this case are $\sim$ 20-30 $\mu$m and 20-40 nm respectively (\textbf{Figure 2(c)}). We also note from FESEM and TEM images (Suppl. Info. \textbf{Figure S1-S3}) that the number of oxide NPs adhering outside CNT is slightly larger in the case of Fe$_{3}$O$_{4}$@CNT as compared to Co$_{3}$O$_{4}$@CNT. However, some big chunks of oxide are additionally observed to co-exist in case of Co$_{3}$O$_{4}$@CNT (Suppl. Info. \textbf{Figure S3}). The inset in \textbf{Figure 2(b)} shows the FESEM image for NiO@CNT, depicting that the sample consists of long and curled CNT in \textit{entangled} morphology. The overall morphology is granular in nature and not carpet like (\textbf{Figure 2(b)}), and the typical length of individual CNT in this case ranges from 1-10 $\mu$m. The filling fraction as well as the residue particles adhering to the CNT are relatively less in NiO@CNT, as compared to Fe$_{3}$O$_{4}$@CNT. It is to be noted that Ni@CNT (and therefore NiO@CNT) is relatively difficult to form in \textit{aligned-forest} morphology, shown in \textbf{Figure 1(a)}.\cite{Kapoor}. It is also to be emphasized  that for all three type of morphologies shown in the inset of \textbf{Figure 2(a-c)},  the well-formed graphitic shells  of CNT  along with  the crystalline oxide - encapsulate  are consistently observed in TEM images \cite{Kapoor}. Such well-formed  and filled CNT were observed for all three samples, either  in \textit{aligned-forest}   (insets of \textbf{Figure 2(c)}) or in \textit{entangled} morphology (\textbf{Figure 2(b)}).

The samples are further characterized using XRD and Raman spectroscopy and a good match with literature was observed. For instance, Raman peaks corresponding  to the carbon nanotubes were observed at $\sim$ 1356 cm$^{-1}$ (D band), $\sim$ 1578 cm$^{-1}$ (G band), and $\sim$ 2715 cm$^{-1}$ (2D or G' band) in case of Fe$_{3}$O$_{4}$@CNT as shown in\textbf{Figure 2(d)}. The D band is assigned to the disordered structures in the hexagonal sp$^2$ carbon network, and the G-band originates from the \textit{in-plane} bond stretching motion of sp$^2$ carbon atoms.\cite{Satio} The increase in the  number of defects on the CNT surfaces may also have contribution from  strain associated with oxide filling.\cite{Kapoor2} The contribution from the  Fe$_{3}$O$_{4}$ is also identified in Raman data.\cite{Faria} The other two samples shown in \textbf{Figure 2(e-f)} were also characterized along similar lines for identification of CNT and the respective oxide-encapsulate.\cite{Ulmane,Hadjiev} \textbf{Figure 2(g-i)} shows the XRD pattern of the Oxides@CNT. The diffraction peak corresponding to CNT is observed at around 26.3$^o$ corresponding to the (002) plane (ICSD code: 015840); and the other peaks were identified as reflections from the corresponding oxides (Fe$_3$O$_4$ JCPDS No.: 75-0033 ; NiO JCPDS No.: 47-1049; Co$_3$O$_4$ JCPDS No.: 42-1467). 

\subsection {Fe$_{3}$O$_{4}$@CNT, Co$_{3}$O$_{4}$ and NiO@CNT: Electrochemical Characterization }

The reversibility and the kinetics of Li ion intercalation and de-intercalation is studied using cyclic voltammetry (CV) measurements. The CV curves for the first three cycles are shown in \textbf{Figure 2(j-l)} for all three Oxides@CNT. In the first cycle shown in \textbf{Figure 2(j)}, a well-defined reduction peak at 0.85 V corresponds to the conversion of Fe$_{2}$O$_3$ into metallic Fe and the formation of SEI film.\cite{Yan,Wang2,Liu1} The anodic peaks at 1.62 and 1.84 V represent the oxidation of Fe to Fe$_{3}$O$_{4}$. The anodic peaks show no significant difference in subsequent cycles, indicating reversibility and capacity stability. Similar data obtained on NiO@CNT and Co$_{3}$O$_{4}$@CNT shown in \textbf{Figure 2(k-l)} are consistent with the previous reports on the electrochemical reaction of the respective bare oxides.\cite{Xu,Ogale} The CV curves are found to be highly repeatable indicating good reversibility for all three types of Oxides@CNT. The electrochemical reaction equation (Suppl. Info.: \textbf{Text S1}) and galvanostatic discharge/charge voltage profiles corresponding to each type of Oxides@CNT are given as (Suppl. Info.:\textbf{ Figure S4(a-c)}).
	
Cyclic stability for sample Fe$_{3}$O$_{4}$@CNT shown in \textbf{Figure 3(a)} at a current density of 200 mA g$ ^{-1} $ scanned for up to 100 cycles. The discharge capacity obtained in the first cycle is 950 mA h g$ ^{-1} $. This delivers a reversible capacity retaining a value of 570 mA h g$ ^{-1} $. The capacity is also observed to  increase slightly during cycling. This tendency has been previously observed for metal oxide electrodes in long-term cycling.\cite{Luo} Although no consensus has been reached in this regard, various possible reasons have been proposed, such as increase in the surface area of the electrode due to pulverization, increase in the conductivity owing to the formation of metallic nanoparticles, etc.\cite{Luo} In our case, this feature appears to correlate  with the \textit{entangled} morphology of Oxide@CNT, in addition to increase in conductivity of the sample. As mentioned previously, the increase in the surface area and secondary voids present in \textit{entangled} CNT can be a contributing factor for its better electrochemical performance. The specific capacity as a function of cycle number for other two samples, NiO@CNT and Co$_{3}$O$_{4}$@CNT is shown in \textbf{Figure 3(b-c)} respectively. Despite the differences in filling fraction of the oxide and the over-all morphology, it is clear that Oxides@CNT exhibit superior cyclic stability. The long-term cycling performance of Oxides@CNT as also tested at higher rate as shown in Suppl. Info. \textbf{Figure} \textbf{S5}. The sample delivers cyclic stability and exhibits reversible capacity of 450 mA h g$ ^{-1} $ up to 600 cycles.

In addition to the cyclic stability, high-rate proficiency of the electrode material is also an important factor for high power applications. The rate performance  is shown in \textbf{Figure 3(d-f)}, and all samples were observed to deliver an outstanding rate performance even at high current densities. In case of Fe$_{3}$O$_{4}$@CNT,  as the current densities are increased to 0.1, 0.25, 0.5, 1, 1.5, 2 A g$^{-1}$, the discharge capacities observed are 515, 472, 447, 425, 410, 390 mA h g$ ^{-1} $ respectively. When the current density is switched to 0.5 A g$^{-1}$, the capacity of Fe$_{3}$O$_{4}$@CNT is restored to a stable capacity of 540 mA h g$^{-1}$. A similar pattern is also observed for other two samples as depicted in \textbf{Figure 3(e-f)}. 

\begin{figure*}[!t]
	\centering
	\includegraphics[width=1\textwidth]{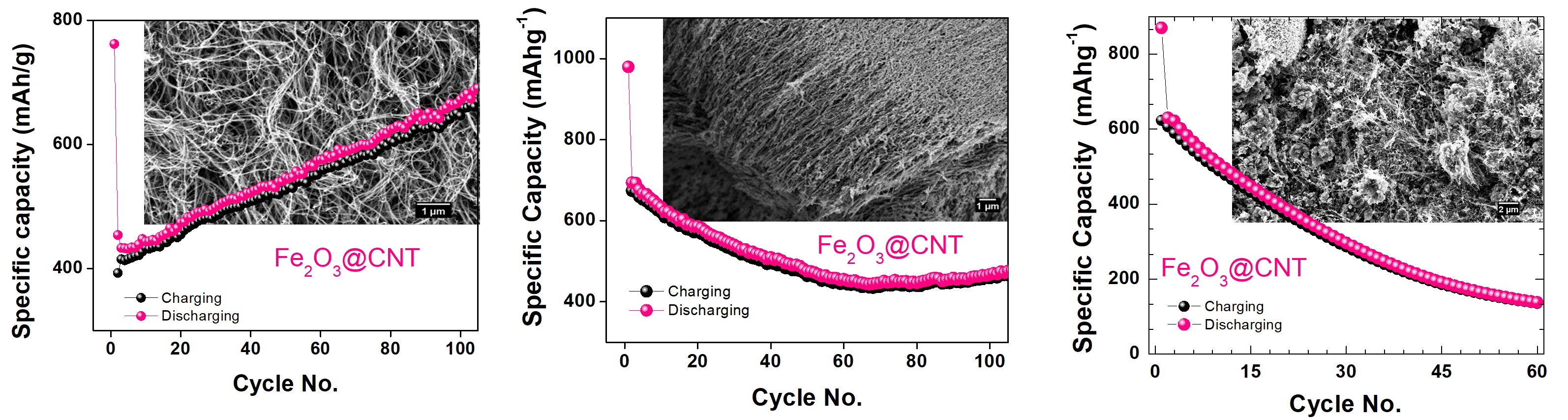}
	\caption{\textit{Effect of morphology and the residue particle density on the cyclic stability of (a) Fe$_{2}$O$_3$@CNT in \textit{entangled} morphology (b) Fe$_{2}$O$_3$@CNT in \textit{aligned-forests} morphology. The morphology of each type of Fe$_{2}$O$_3$@CNT sample is depicted in the respective FESEM images shown in the inset. (c) Cycling performance of sample Fe$_{2}$O$_3$@CNT with large number of residue oxide nanoparticles outside CNT conducted using a Swagelok-type cell. The morphology of the sample is depicted by the FESEM shown in the inset.The filling efficiency is also low is this case.}}
	\label{fig:4}
\end{figure*}

It is interesting to note that all the three Oxides@CNT not only exhibit exceptional cyclic stability up to 100 cycles and for discharge rate 100 mA/g  but also exhibit reasonable capacity, which is $\sim$ 500-600 mA h g$^{-1}$. Here the magnitude of capacity is highest for Fe$_{3}$O$_{4}$@CNT which has a higher oxide content than NiO@CNT. However, Fe$_{3}$O$_{4}$@CNT forms in a carpet like structure, which may be less conducive for Li-ion transportation as the filled CNT residing deep within the carpet may not be available as an active material as opposed to, say NiO@CNT, which has been formed in an \textit{entangled} structure. 

For further exploration of the correlation between morphology and the magnitude of the capacity, we performed Electrochemical Impedance Spectroscopy (EIS) measurements for all three samples. The corresponding Nyquist plot obtained is shown in \textbf{Figure 3(g-i)}. On a general note, the large semicircle in the mid-frequency region of the Nyquist plot corresponds to the charge transfer resistance (R$_{ct}$), and the straight line in the low-frequency region characterizes the Warburg impedance of the Li ion diffusion.\cite{Chen} The region of semicircle (corresponding to higher frequencies) in the Nyquist plot typically gives information about electrode resistance, reaction rate and double layer capacitance etc., whereas the straight line (lower frequency region) reflects the diffusion coefficient of ions. Though a detailed analysis of Nyquist plot for each Oxide@CNT sample is beyond the scope of the present work, we focused on the semi-circular arc in the Nyquist plot for a comparative analysis of all three type of Oxides@CNT. This region contains information about variations in the electrical resistance due to specific morphology and also to an extent depends on the electrode preparation details

From \textbf{Figure 3(g)} we note that the diameter of the semicircle in the mid-frequency region is significantly smaller than what has been observed in the previous reports on iron-oxide electrode materials.\cite{Yan,Huang,Liu1} The low Ohmic resistance signals easy electron transfer during the electrochemical Li ion insertion-extraction, resulting in an enhanced electrochemical performance of the electrode material. The R$_{ct}$ for Fe$_{3}$O$_{4}$@CNT is found to be $\sim$ 28 $\Omega$ as is evident from \textbf{Figure 3(g)}. The lower R$_{ct}$ is presumably due to an increased electrical conductivity facilitated by the well-connected CNT network in the carpet-like structure. Similar data on other two oxides reveal that R$_{ct}$ for NiO@CNT and CoO$_x$@CNT is $\sim$ 100 $\Omega$ and $\sim$ 200 $\Omega$ respectively (\textbf{Figure 3(h-i)}). We also note that Fe$_{3}$O$_{4}$@CNT and Co$_{3}$O$_{4}$@CNT exhibit similar value of R$_{ct}$ when a fresh cell is compared with the cell after five cycles. On the other hand, R$_{ct}$ for a fresh cell is significantly different from what is observed after five cycles in the case of NiO@CNT.

From the FESEM images recorded at higher magnification in the inset of \textbf{Figure 3(g-i)}, we note that NiO@CNT and Fe$_{3}$O$_{4}$@CNT form in \textit{entangled} morphology, but the CNT network connectivity is evidently better for Fe$_{3}$O$_{4}$@CNT which has a carpet-like structure (inset of \textbf{Figure 2(a)}). On a similar note, the CNT network connectivity is also good for Co$_{3}$O$_{4}$@CNT. While an \textit{aligned-forest} structure may be less conducive for Li ion transport, the better CNT network connectivity of the electrode in this case is also likely to result in R$_{ct}$ after five cycles to be similar to the fresh cell. NiO@CNT exhibit long and curled CNT in \textit{entangled} morphology, but overall the CNT network is not uniform. It forms a rather granular structure, as is evident from the FESEM image shown in the inset and in Suppl. Info. \textbf{Figure S2}. 

Regarding the magnitude of capacity, we note that in the case of Co$_{3}$O$_{4}$@CNT, TGA data have shown 54\% oxide content. However, multiple FESEM images reveal the presence of big  chunks of isolated oxide particles in this sample (Suppl. Info. \textbf{Figure S3}). This is in addition to the oxide NPs adhering outside the CNT. The filling efficiency in this case is certainly lower than Fe$_{3}$O$_{4}$@CNT. The big chunks of oxide particles adversely affect the overall battery performance in the case of Co$_{3}$O$_{4}$@CNT. (Roughly the filling efficiency for all three samples is in the range of 20 to 25\%, wherein relatively higher filling efficiency is observed for FeO$_x$@CNT.\cite{Kapoor})    

From the data obtained so far, it is evident that Oxides@CNT lead to superior cyclic stability. Moreover, it is also clear that factors such as morphology, and presence of residual oxide NPs appear to crucially influence the overall battery performance. We also observed that electrode preparation details, including sonication of powders prior to cell fabrication, and connectivity of the CNT network appear to influence the battery performance. To further investigate all these factors we tested Fe$_{2}$O$_{3}$@CNT in three different morphologies.

\subsection{Fe$_{2}$O$_3$@CNT : Effect of Morphology on Electrochemical Data}

\textbf{Figure 4} compares the cyclic performance of Fe$_{2}$O$_{3}$@CNT, in three different morphologies, for which FESEM images are shown in the respective inset. The samples with \textit{entangled} and \textit{aligned-forest} morphologies in \textbf{Figure 4(a-b)} are tested for LIBs using a coin cell. Both the samples shown in \textbf{Figure 4(a-b)} contain residual oxide NP, which are similar in number density. In addition, another Fe$_{2}$O$_{3}$@CNT sample with much larger fraction of residual NPs (as compared to the ones shows in \textbf{Figure 4(a-b)}) is also investigated. Morphology of this sample can be seen in the FESEM image in the inset of \textbf{Figure 4(c)}. The overall filling efficiency is lower in this case, as compared to the samples shown in \textbf{Figure 4(a-b)}. XRD data (not shown here) revealed that the sample also contains a tiny amount of Fe$_{3}$O$_{4}$. 

It is interesting to note that the sample shown in \textbf{Figure 4(a)}, consisting of long \textit{entangled} CNT, exhibits an increasing trend in the magnitude of specific capacity as a function of the cycle number. The FESEM image shown in \textbf{Figure 4(a)} reveals that long and curled CNT are uniformly formed in \textit{entangled} morphology. As is evident from the main panel, an initial capacity 400 mA h g$^{-1}$ increases up to 700 mA h g$^{-1}$ after 100 cycles. This type of trend has been observed earlier for Fe$_{3}$O$_{4}$@CNT in \textit{entangled} morphology, but with carpet-like structure \textbf{Figure 2(a)}). However, \textit{entangled} morphology with a uniform texture in the case of Fe$_{2}$O$_{3}$@CNT leads to an increase in capacity, measured for about 100 cycles. The \textit{aligned-forest} structure, on the other hand, shows a slight tendency for reduction in capacity over cycle number, as is evident from \textbf{Figure 4(b)}. Thus, the results suggest that the \textit{entangled} morphology with long and curled tube, well-connected CNT network and reasonable filling efficiency  is more favorable for better cycle performance. The results of Fe$_{2}$O$_{3}$@CNT sample (for which Swagelok-type of cell was used for characterization), with a large number of oxide NPs, and with less filling efficiency, exhibit a faster capacity fade as shown in the main panel of \textbf{Figure 4(c)}.  

We also compared the same batch of Fe$_{3}$O$_{4}$@CNT sample, tested using a coin cell and Swagelok-type cell. For both the measurements, the Fe$_{3}$O$_{4}$@CNT powders were also sonicated prior to the electrode preparation. These results are shown in \textbf{Figure S6}. Here the magnitude of capacity is about 450 mA h g$^{-1}$ with good cyclic stability observed in both the cases. However, the minor differences can be attributed to electrode preparation details and the final texture of the pressed powders in both cases. Here a slight difference can  also arise due to the changes in conductance of the pressed powders. Overall, we conclude that the encapsulation of oxide within the core cavity leads to better cyclic stability. The oxide NPs or isolated chunk of bare oxide may add to the magnitude of specific capacity but their presence is detrimental to cyclic stability, as is evident from \textbf{Figure 4}. Thus, superior cyclic stability is primarily connected to encapsulation of oxide within the core cavity of the CNT.  

\subsection{Oxides@CNT: Superior Cyclic Stability}

\begin{figure}[!t]
	\centering
	\includegraphics[width=0.5\textwidth]{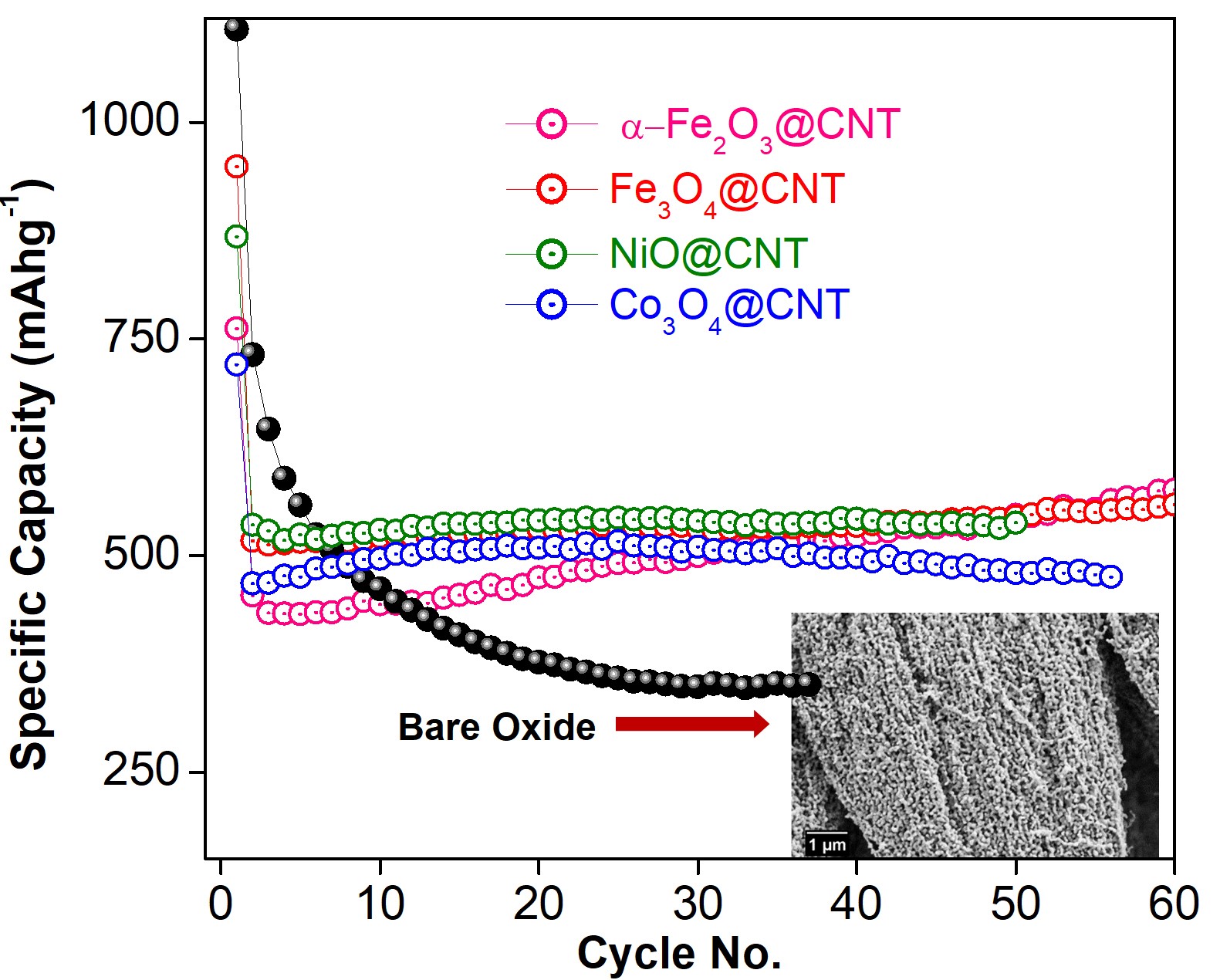}
	\caption{\textit{Oxides@CNT (coloured dots) deliver a stable discharge capacity, irrespective of the type of the oxide encapsulate, as compared to the bare oxide (black dots) which exhibits a decaying trend after the initial cycles. Here the bare oxide is Fe$_{2}$O$_3$ in the form of \textit{aligned-forests}, for which FESEM is shown in the lower inset.}}
	\label{fig:5}
\end{figure}

To confirm that superior cyclic stability in these Oxides@CNT nano-structures arises from encapsulation within the core cavity of the CNT, we  conducted control  measurements on the bare-oxide. Here the representative bare-oxide ($\alpha$-Fe$_{2}$O$_3$) is formed in the same morphology as Fe$_{2}$O$_3$@CNT by suitable annealing \cite{Kapoor2}.   The bare-oxide template was obtained from the same batch of Fe$_{2}$O$_3$@CNT, on which electrochemical measurements were performed. This enables us to check and isolate the effect of the morphology of the active material on the cyclic stability. These data, in conjunction with all Oxides@CNT are presented in \textbf{Figure 5}. The morphology of the representative bare-oxide is shown in the inset of Figure 5. The morphologies of other Oxides@CNT are shown in the inset of \textbf{Figure 2 and Figure 4(b)}.

The black dots in \textbf{Figure 5} display the cycling performance of bare oxide template, which is Fe$_{2}$O$_3$ at a current density of 100 mA g$^{-1}$. The specific capacity of bare oxide template drops down from 1100mA h g$^{-1}$ at first cycle to 350 mA h g$^{-1}$ after 35 cycles, thus, exhibiting a poor stability.\cite{Ji} The primary reason for poor cyclic stability of the bare oxide materials in LIBs is known to arise from pulverization of the anode material. The drastic volume variations during metal to metal-oxide conversions leads to strain effects during the cyclic process, which together with low electrical conductivity in bare oxide leads to poor cycle life\cite{Ji}.  Fe$_{2}$O$_3$@CNT, on the other hand exhibits superior cyclic stability (pink dots in \textbf{Figure 5}) and cycle capacity retention. It is also evident that bare oxide ($\alpha$-Fe$_{2}$O$_3$ in this case) exhibit poor cyclic stability as compared to not only Fe$_{2}$O$_3$@CNT but all other Oxides@CNT, presented in \textbf{Figure 5}. These oxides@CNT contain  different type of oxide encapsulate with varying  filling fractions  and  density of residue oxide NP. More importantly, data is on Oxides@CNT samples formed in different types of morphologies, such as \textit{aligned-forests} or \textit{entangled}. Nevertheless, all these oxides@CNT exhibit superior cyclic stability as compared to the bare-oxides. Thus, encapsulation of oxides within the core cavity of CNT accommodates strain related to metal to metal-oxide conversion.\cite{Yan} It appears that factors such as filling fraction, residue oxide NP and overall morphology of the oxide@CNT can play an important role in tuning the magnitude of specific capacity. Some of these factors need optimization: for instance, the \textit{entangled} morphology may be more conducive to Li-ion transportation, but the \textit{aligned-forests} morphology, owing to better connectivity of the CNT network leads to improved conductance. 

\section{Conclusions}
We have conducted comprehensive electrochemical measurements on samples of carbon nanotubes, in which four different type of transition metal oxides have been encapsulated. This includes Fe$_{3}$O$_4$@CNT, Fe$_{2}$O$_3$@CNT, NiO@CNT, and Co$_{3}$O$_4$@CNT; all of which have been synthesized in desired morphology to optimize best parameters for LIBs. The samples have been characterized using X-Ray diffraction, Raman, Thermogravimetric analysis, Scanning and Transmission Electron Microscopy. These characterization tools enable us to correlate the structural aspects including morphology of the CNT network, filling efficiency, and the residue particle density with the electrochemical data. The electrochemical performance is investigated using a coin cell as well as a Swagelok-type of cell configurations. Our data enables us to conclude that Oxides@CNT deliver a superior cyclic stability as compared to bare-oxides. The excellent stability and high capacity of Oxides@CNT can be attributed to the introduction of carbon nanotube structures that contain the oxide within its core cavity. The graphitic shells of the CNT buffer the volumetric strain effects that take place during metal to metal-oxide conversion and improves the electrical conductivity, consequently resulting in excellent cycling performance. The CNT also protects the electrode materials from pulverization, which in-turn offers long-term stability of the electrode materials. We observe that the morphology of the CNT network also plays an important role. The \textit{entangled} morphology of the CNT network with reasonable filling efficiency and control on the residue oxide nano-particles  leads to good cyclic stability as well as higher capacity.  This systematic study enables optimization of best parameters for the usage of these synthetically encapsulated and self-organized Oxides@CNT nano-structures for electrochemical applications.

\section*{Acknowledgments}
The authors thank Mr. J. Parmar, Mr. S.C. Purandare and Mr. R. Bapat (TIFR) for TEM measurements; Mr. Anil Shetty (IISER-P) for SEM, Miss Edna Joseph (NCL, Pune) for TGA measurements. The authors also thank F. Wilhelmi and L. Deeg for experimental support. AB acknowledges DST, India for funding support through a Ramanujan Grant. SO and AB acknowledge DST Nano mission Thematic Unit Grant.

\section*{Competing Interests}
The authors declare no competing interests.

\bibliographystyle{apsrev4-1}
\bibliography{Ref}

\end{document}